\documentclass[sigconf, natbib, screen]{acmart}
\pdfoutput=1 
\AtBeginDocument{%
  \providecommand\BibTeX{{%
    \normalfont B\kern-0.5em{\scshape i\kern-0.25em b}\kern-0.8em\TeX}}}

\setcopyright{acmcopyright}
\copyrightyear{2023}
\acmYear{2023}
\acmDOI{10.1145/TBA}

\acmConference[SIGCSE '23]{SIGCSE '23: ACM Technical Symposium on Computer Science Education}{March 15--18, 2023}{Toronto, Canada}
\acmBooktitle{SIGCSE '23: ACM Technical Symposium on Computer Science Education,
  March 15--85, 2023, Toronto, Canada}
\acmPrice{15.00}
\acmISBN{978-1-4503-XXXX-X/18/06}

\usepackage{lipsum} 
\usepackage[main=english]{babel} 

\usepackage{xcolor} 
\begin{document}

\title{Integrating Accessibility in a Mobile App Development Course}

\author{Jaskaran Singh Bhatia}\affiliation{%
\institution{BITS Pilani, KK Birla Goa Campus}
\city{}\state{Goa}
\country{India}
\postcode{403726}
}
\author{Parthasarathy P D}\affiliation{%
\institution{BITS Pilani, KK Birla Goa Campus}
\city{}\state{Goa}
\country{India}
\postcode{403726}
}
\author{Snigdha Tiwari}\affiliation{%
\institution{BITS Pilani, KK Birla Goa Campus}
\city{}\state{Goa}
\country{India}
\postcode{403726}
}
\author{Dhruv Nagpal}\affiliation{%
\institution{BITS Pilani, KK Birla Goa Campus}
\city{}
\state{Goa}
\country{India}
\postcode{403726}
}
\author{Swaroop Joshi}\email{swaroopj@goa.bits-pilani.ac.in}\orcid{0000-0003-4536-2446}
\affiliation{%
  \institution{BITS Pilani, KK Birla Goa Campus}
  \city{}\state{Goa}
  \country{India}
  \postcode{403726}
}


\begin{abstract}
The growing interest in accessible software reflects in computing educators' and education researchers' efforts to include accessibility in core computing education. 
We integrated accessibility in a junior/senior-level Android app development course at a large private university in India. 
The course introduced three accessibility-related topics using various interventions: Accessibility Awareness (a guest lecture by a legal expert), Technical Knowledge (lectures on Android accessibility guidelines and testing practices and graded components for implementing accessibility in programming assignments), and Empathy (an activity that required students to blindfold themselves and interact with their phones using a screen-reader).
We evaluated their impact on student learning using three instruments: (A) A pre/post-course questionnaire, (B) Reflective questions on each of the four programming assignments, and (C) Midterm and Final exam questions. 
Our findings demonstrate that: (A) significantly more ($p<.05$) students considered disabilities when designing an app after taking this course, 
(B) many students developed empathy towards the challenges persons with disabilities face while using inaccessible apps, and 
(C) all students could correctly identify at least one accessibility issue in the user interface of a real-world app given its screenshot and 90\% of them could provide a correct solution to fix it.
\end{abstract}



\maketitle

\newcommand{\tb}{Talk Back}

\newcommand{\as}{Accessibility Scanner}

\newcommand{\esp}{Espresso}

\raggedbottom

\section{Introduction}
\label{sec:intro}

Despite an increasing awareness in the software industry about accessibility over the past few years, many mainstream software products fall short of meeting accessibility criteria~\citep{rossEpidemiologyinspiredLargescaleAnalysis2020, agrawalAssessingUsabilityAccessibility2021, fokLargeScaleLongitudinalAnalysis2022, balajiAccessibilityAnalysisEgovernance2016, kuppusamyEvaluatingWebAccessibility2021}.
Software professionals point to a lack of relevant knowledge and skills in computing education for this gap in implementing accessibility~\citep{patelWhySoftwareNot2020}. 
According to the computing (CS) faculty, the main barriers to teaching accessibility include the lack of clear learning objectives about accessibility and their lack of knowledge about accessibility~\citep{shinoharaWhoTeachesAccessibility2018, soaresguedesHowAreWe2020}. 
Also, most instructors who teach accessibility have a background in human-computer interaction (HCI) and related fields and teach these topics in specialized electives~\citep{shinoharaWhoTeachesAccessibility2018}. 
This indicates that other instructors may find it challenging to build their expertise and incorporate accessibility topics in the ``mainstream'' CS courses.
Our work contributes to the literature that aims to bridge this gap by developing relevant course material, resources, and assignments that other instructors can readily integrate into their courses without much prior knowledge about accessibility and without diluting the ``core'' learning objectives.

We report on our experience integrating accessibility topics in a mobile app development elective for junior/senior undergraduate CS students. 
All resources, including slides, video lectures, assignments, starter codes, and exam questions, are available online at \url{https://swaroopjoshi.in/project/sugamyata/}. 
We describe the course (Sec.~\ref{sec:course}), the instruments we used to evaluate the effects of integrating these topics and our findings (Sec.~\ref{sec:methods-findings}), and discuss the results (Sec.~\ref{sec:discussion}) and directions for future work (Sec.~\ref{sec:future}) in the rest of this paper.

\section{Related Work}
\label{sec:bg}


Educators have reported on teaching accessibility in special topics courses such as assistive technology~\citep{matauschAssistecUniversityCourse2006} and universal access~\citep{kurniawanGeneralEducationCourse2010} and developing specialized modules for graduate programs~\citep{wald2008design}. 
Some others have integrated accessibility topics in existing courses such as web development~\citep{rosmaitaAccessibilityFirstNew2006,freireAccessibilityWebMultimedia2013,wangHolisticPragmaticApproach2012}, artificial intelligence~\citep{tsengExplorationIntegratingAccessibility2022}, software engineering~\citep{ludiIntroducingAccessibilityRequirements2007, el-glalyTeachingAccessibilitySoftware2020, rossAccessibilityFirstclassConcern2017}, HCI~\citep{palanTeachingInclusiveThinking2017,poorNoUserLeft2012}, and introduction to  programming (CS1/CS2)~\citep{jiaInfusingAccessibilityProgramming2021,rossAccessibilityFirstclassConcern2017,cohenAccessibilityIntroductoryComputer2005}.
Such courses mainly focus on teaching: (i) Accessibility awareness, (ii) Technical knowledge like tools for accessibility testing, (iii) Empathy, and (iv) Potential endeavours using various instructional methods such as in-class activities, projects, lectures, assignments, videos, simulated disabilities, interaction with people with disabilities, guest lectures, and research~\citep{bakerSystematicAnalysisAccessibility2020}. Other instructor resources include materials by the AccessComputing initiative~\citep{koAccessComputingPromotesTeaching2016}, Accessibility guidelines like WCAG 2.0~\citep{bencaldwellWebContentAccessibility2008}, Accessibility Learning Labs~\citep{el-glalyPresentingEvaluatingImpact2020}, the Mobile inclusive learning kit~\citep{el-glalyAppsEveryoneMobile2018}, and games for accessibility awareness~\citep{kletenikLetPlayIncreasing2022}.



Our work differs from the existing literature in that, to our knowledge, this is the first report on a mobile app development course with accessibility as an underlying theme~\citep{bakerSystematicAnalysisAccessibility2020}, with multiple assignments and evaluative components assessing the accessibility knowledge of students throughout the semester.
Another contribution is the Inclusive Thinking Questionnaire, which can be used in other contexts to assess whether the respondents consider various inclusivity criteria when designing software.

\section{The Course}
\label{sec:course}

We report on the Fall 2021 offering of a junior/senior-level Mobile App Development elective at a private university in India. This course focuses on Android app development using Java. Most of the students were CS majors with prior coursework in Object-oriented programming, UML, and software engineering fundamentals.
One of the stated learning objectives was ``Understand the importance of developing accessible software and demonstrate integrating accessibility components in Android apps.''
The course was conducted  online due to the COVID-19 pandemic via the Google Meet platform. The class was taught in a ``flipped'' mode: pre-recorded video lectures were made available to the students before the class, and the class meeting hours were used for several in-class active-learning activities. 
It covered the basics of mobile app development and various Android-specific topics such as activity lifecycle, fragments, navigation components, services, background tasks, data persistence, permissions, and sensors. It also introduced cross-platform development using Flutter towards the end.
The course had four programming assignments worth 30\% of the final grades and two exams worth 25\% and 30\%, respectively. The remaining 15\% were for in-class activities.
One of the authors of this work was the instructor, and another author was an undergraduate teaching assistant of the course.

\subsection{Interventions}
\label{sec:method:interventions}

While the primary focus of the course was on software engineering concepts and Android features, we introduced the following interventions to incorporate Accessibility Awareness, Accessibility Knowledge, and Empathy towards persons with disabilities among the students.

\subsubsection{Guest Lectures}

To raise awareness about the existing laws about disabilities and expose students to the challenges persons with disabilities may face in their day-to-day lives, a guest lecture by a lawyer with expertise in the field was organized. He talked about the Rights of Persons with Disabilities (RPD), 2016 act of the Indian Parliament~\citep{ministryofsocialjusticeandempowermentgovernmentofindiaRightsPersonsDisabilities2016}, recounted his work on the relevant cases, and shared his thoughts on how computing graduates can make an impact in the field of accessibility. He answered several student questions at the end.


\subsubsection{Classroom Lectures}

Accessibility was introduced early on in the course. In the third week of classes, after having introduced the basics of Android app development and the fundamentals of testing, the instructor talked about the statistics on disabilities and the need for developing accessible software. 
Next, the instructor presented the Android Accessibility Guidelines~\footnote{\url{https://developer.android.com/guide/topics/ui/accessibility}} and demonstrated \tb{} and \as{}, two softwares from the Android Accessibility Suite to evaluate the accessibility features of an app under development. 
The former is a screen reader developers can use to check how their app \emph{feels} to a visually impaired user, and the latter is a tool that scans the UI of an app and produces reports on its compliance with certain accessibility guidelines like the minimum size of interactive widgets. 
Later, methods for running accessibility checks via automated UI testing using the \esp{} framework were introduced.

\subsubsection{Programming Assignments}

Four programming assignments spread across the semester assessed how well students implemented various features of Android app development. In addition, we added some accessibility-related tasks to assess students' accessibility knowledge (See Sec.~\ref{sec:method:assignments}).


\subsection{Participants}

The course instructor, who is also a co-author of this paper, explained the purpose of the research to the students and shared the informed consent form approved by the institute’s Human Ethical Committee (HEC) with them. 50 out of the 72 enrolled students consented to participate in the study. Data from these students were anonymized and handed over to the researchers over secure cloud storage for analysis. All participants were junior or senior year undergraduate students aged 18 to 22. Forty-seven of them were males, and three were females.

\section{Methods and Findings}
\label{sec:methods-findings}
We used the following three instruments to measure the impact on Accessibility Awareness, Technical Skills, and Empathy of the students:

\begin{description}
  \item[A] An Inclusive Thinking Questionnaire at the start and end of the course,
  \item[B] Reflective questions on accessibility as part of the programming assignments, and
  \item[C] Questions on accessibility in both the exams.
\end{description}

We discuss each in detail below.

\subsection{Inclusive Thinking Questionnaire (A)}

This instrument, adapted from Ludi's voting kiosk scenario~\cite{ludiIntroducingAccessibilityRequirements2007, palanTeachingInclusiveThinking2017}, was used to understand the change in Accessibility Awareness of the students due to exposure to the interventions in the course. The questionnaire asked the participants to design a hypothetical ``COVID-19 vaccination verification'' and posed three open-ended questions about the design and testing of the app prototype.
The text of the questionnaire is as follows:

\begin{quote}
  Let us consider a hypothetical scenario: After the successful COVID-19 vaccination of most adults, the government has decided to relax all restrictions -- no lockdowns anymore! But, for almost everything that requires interaction with someone outside your household, one may be asked to present proof of vaccination. You are hired to design a software solution that enables (a) producing the proof of vaccination when asked and (b) verifying a `proof'. Remember, anyone and everyone should be able to do both these things -- a shopkeeper can ask a customer for proof and deny entry if it's not verified, for example. 
  
  Now, answer these three questions to the best of your abilities and in as much detail as possible.
  \begin{enumerate}
    \item Discuss some key points of the design of this system (interfaces, input-output, etc.).
    \item What potential challenges do you see in the large-scale adoption of this solution?
    \item Who will be your potential users for testing your prototype to gain feedback on the design?
  \end{enumerate}
\end{quote}

The first and the third question are the design and evaluation questions from Ludi's questionnaire. The second question was added to nudge the participants to think about various inclusivity issues in the large-scale adaption of the system. 
%
The questionnaire was first administered at the start of the course, even before any accessibility topics were mentioned, and again in the week before the final exams.

\subsubsection{Data and Analysis}

The data was collected via a google form at the start and the end of the semester. After removing responses from participants who did not complete both the pre- and post-course questionnaire, two coders analyzed data from 40 participants.
We followed the methods of Salda\~na~\cite[pp.~115--121]{saldanaCodingManualQualitative2021} to perform a Two-pass Mangnitude Coding on this data. Two researchers independently identified codes from 10\% of the sample and agreed on the definitions of the codes before classifying the rest of the data using these codes:
\begin{itemize}
  \item \texttt{INFRA}: considers the infrastructural limitations, for instance, the lack of availability of high-speed internet in some areas.
  \item \texttt{TECH}: considers the limitations due to lack of affordability or comfort using technology.
  \item \texttt{LANG}: considers the language barrier, particularly the lack of English proficiency.
  \item \texttt{LITERACY}: considers making the proposed solution usable by people with varying degrees of basic literacy.
  \item \texttt{AGE}: considers older people when designing or testing the solution.
  \item \texttt{IMPL-DIV}: considers various diversities in society, even implicitly, for instance, by stating that the testing should be done with various socio-economic groups.
  \item \texttt{DIS}: considers persons with disabilities when designing or testing the solution.
\end{itemize}

The codes were then synthesized into these three categories:
\begin{itemize}
  \item \textbf{Infrastructure Barriers}: \texttt{INFRA}
  \item \textbf{Diversity}: \texttt{TECH}, \texttt{LANG}, \texttt{LITERACY}, \texttt{AGE}, \texttt{IMPL-DIV}
  \item \textbf{Accessibility}: \texttt{DIS}
\end{itemize}

\subsubsection{Findings} 

\begin{table*}[htbp]
  \centering
\begin{tabular}{|l|rr|rrrr|c|}
  \hline
  Category                  &  Pre  & Post & a & b & c & d & $p$ \\\hline
  \textbf{Infrastructure Barriers}&  10   & 17   &4&13&6&17& .168668\\  
  \textbf{Diversity}              &  25   & 27   &19&8&6&7& .789268\\
  \textbf{Accessibility}          &  1    & 17   &1&16&0&23& .000177\\\hline
\end{tabular}
\caption{Categorywise responses to the Inclusive Thinking Questionnaire ($N=40$) with $p$-values using McNemar's test. $a$ and $d$ are the concordant cells and $b$ and $c$ are the discordant cells.}\label{tab:questionnaires}
\end{table*}

Table~\ref{tab:questionnaires} shows the results for each category in the pre- and post-course questionnaire ($N=40$). 
Columns $a$ and $d$ are concordant: $a$ is the number of participants whose response shows the presence of the codes in a category in both measurements, and $d$ is the number of participants whose response shows the absence of the codes in both. 
$b$ and $c$ are discordant: the former is the number of participants whose response showed the absence of that category in the pre-test but presence in the post-test, while the latter is the opposite. 
To illustrate, only one student considered disabilities in both measurements, while 23 did not consider disabilities in either. Sixteen students did not consider disabilities in the first measurement but did so in the second, and there was no student who considered disabilities in the pre-test but did not consider it in the post-test.


Since we have dichotomous dependent variables in each category ---either the participants `considered' the category or they `did not consider' it--- and we have taken repeated measurements, we use McNemar's test~\cite{pemburysmithEffectiveUseMcNemar2020} to determine if there was a significant difference between the two measurements in each category.
The difference in the accessibility category is statistically significant: 
$p = .000177$. However, no significant difference is observed in the other two categories.

\subsection{Accessibility Evaluation on Programming Assignments (B)}
\label{sec:method:assignments}

Four programming assignments assessed students on various features of Android and software engineering concepts introduced throughout the semester.
Students were provided with a starter code for each assignment via GitHub Classroom. The first three assignments had explicit requirements for accessibility, worth 5, 10, or 15 points out of 100, as shown below. 
The fourth assignment did not have an explicit accessibility-related task as we wanted to check if students implement accessibility even without a `grade incentive'. 
All four assessments had open-ended questions on experiences of using the accessibility tools.


\subsubsection{Assignment 1} Students implemented a ``single-screen app'' that asks the user to enter a date and reports the day of the week (Monday, Tuesday, etc.) for that date. It required some error checking for dates that do not exist (e.g., 31 April). Accessibility tasks were worth 5 points. Reflective questions:
\begin{itemize} 
  \item Run the app with your eyes closed, using \tb{}. Write a paragraph about your experience doing so.
  \item Run \as{} on your app. List all the suggestions it identifies to make your app more accessible (it should find at least one\footnote{The starter code violated some accessibility requirements by design.}). Modify your app so that the next run of \as{} cannot find any suggestions.  Briefly describe how you fixed the problems it first identified. 
\end{itemize}

Learning objectives for the accessibility tasks (with the corresponding Revised Bloom Taxonomy~\citep{krathwohlRevisionBloomTaxonomy2002} verb underlined):
\begin{itemize}
  \item LO1: \underline{Understand} difficulties that a user with visual impairments might encounter when using inaccessible apps.
  \item LO2: \underline{Apply} the assistive technology tools \tb{} and \as{}.
  \item LO3: \underline{Create} a more accessible version of a given app.
\end{itemize}

\subsubsection{Assignment 2} A dice-game that awarded the player points on guessing some properties of the dice throws (e.g., the player gets 5 points if they correctly guess whether the sum of the dice faces after the throw will be an even number).
Accessibility tasks were worth 15\% points on this assignment.
Questions:
\begin{itemize} 
  \item Run the app with your eyes closed, using \tb{}. Write a paragraph about your experience doing so.
  \item This time, you will test your app's accessibility by integrating it with \esp{}. Write at least four \esp{} test cases that interact with different UI elements.
\end{itemize}

Learning objectives for the accessibility tasks:
\begin{itemize}
  \item LO4: \underline{Understand} difficulties that a user with visual impairments might encounter when using inaccessible apps.
  \item LO5: \underline{Apply} the assistive technology tool \tb.
  \item LO6: \underline{Create} automated tests for an Android app in Java using \esp.
  \item LO7: \underline{Analyze} an app using \esp{}.
\end{itemize}

\subsubsection{Assignment 3} A ``Journal'' for recording user's daily activities that uses a local database. Accessibility tasks were worth 10 points.
Questions:
\begin{itemize}
  \item You will again test your app's accessibility by integrating it with \esp{}, as you did in A2. You don't have to write any new code for this, except for the appropriate annotations.
\end{itemize}

Learning objectives for the accessibility tasks:
\begin{itemize}
  \item LO8: \underline{Create} automated tests for an Android app in Java using \esp.
  \item LO9: \underline{Analyze} an app using \esp{}.
\end{itemize}

\subsubsection{Assignment 4} The fourth assignment was a two-player Tic-Tac-Toe game implemented using Google's Firebase as the backend.
There were no explicit tasks for accessibility and no grade points.
Questions:
\begin{itemize}
  \item Did you consider testing and accessibility? Please mention what you did, if you did it. Please mention that you didn't, even if you didn't.
\end{itemize}

\subsubsection{Data and Analysis} For each assignment, in response to the open-ended questions, students self-reported their experience using one or more of the three tools for testing the accessibility of the apps they developed. 
We performed Magnitude Coding on these responses as described in the previous section with these codes:
\begin{itemize}
  \item \texttt{TB}: Accessibility testing was done using the \tb{} tool,
  \item \texttt{AS}: Accessibility testing was done using the \as{} tool,
  \item \texttt{ESP}: Accessibility testing was done using \esp{} instrumented tests, and
  \item \texttt{NONE}: Accessibility testing was not done.
\end{itemize}

\subsubsection{Findings}
\begin{table}[t]
  \centering
  \begin{tabular}{|c|rrrr|}
    \hline
    Code           & A1 & A2  & A3  & A4\\
    \hline
    \texttt{TB}   & 49 & 35  & 7  & 4 \\
    \texttt{AS}   & 49 & 8   & 6  & 8 \\
    \texttt{ESP}  & NA & 33  & 15 & 9 \\
    \texttt{NONE} & 1  & 6   & 35 & 41 \\\hline 
    Points for accessibility tasks & 5 & 15 & 10 & 0\\\hline
  \end{tabular}
  \caption{Accessibility testing in programming assignments ($N=50$). (NA: the topic was not introduced yet)\label{tab:assignments}}
\end{table}

Table~\ref{tab:assignments} shows the number of student responses per code per assignment ($N=50$). Note, \esp{} was not introduced to the students at the time of A1. 

49 and 44 students out of 50 have used the required accessibility testing tools in the first two assignments, respectively. These assignments had explicit accessibility-related tasks: using \tb{} and \as{} in the first assignment and using \tb{} and \esp{} annotated tests in the second. Eight students voluntarily used \as{} in A2 even when it was not a stated requirement.
Similarly, in A3, seven students used \tb{} and six \as{} even though that was not a stated requirement. 
However, the number of students who used \esp{} is only 15 out of 50. We surmise this is because the third assignment had a lot of ``technical'' requirements like designing and implementing a database and using navigation components and menu bar actions in the app. And, with only 10 out of 100 points for accessibility testing, some students might have focused on the other requirements of the app.

In A4, although there was no stated accessibility requirement and hence no points for using any of these tools, nine students reported that they had tested their apps for accessibility, and eight others stated that they had planned to but could not do so due to lack of time.

\subsection{Exam questions (C)}
\label{sec:method:exams}

The midterm and final exams had questions to assess the students' Accessibility Knowledge. 

\subsubsection{Midterm question} 
\begin{figure}[thbp]
  \centering
  \fbox{\includegraphics[width=.4\textwidth]{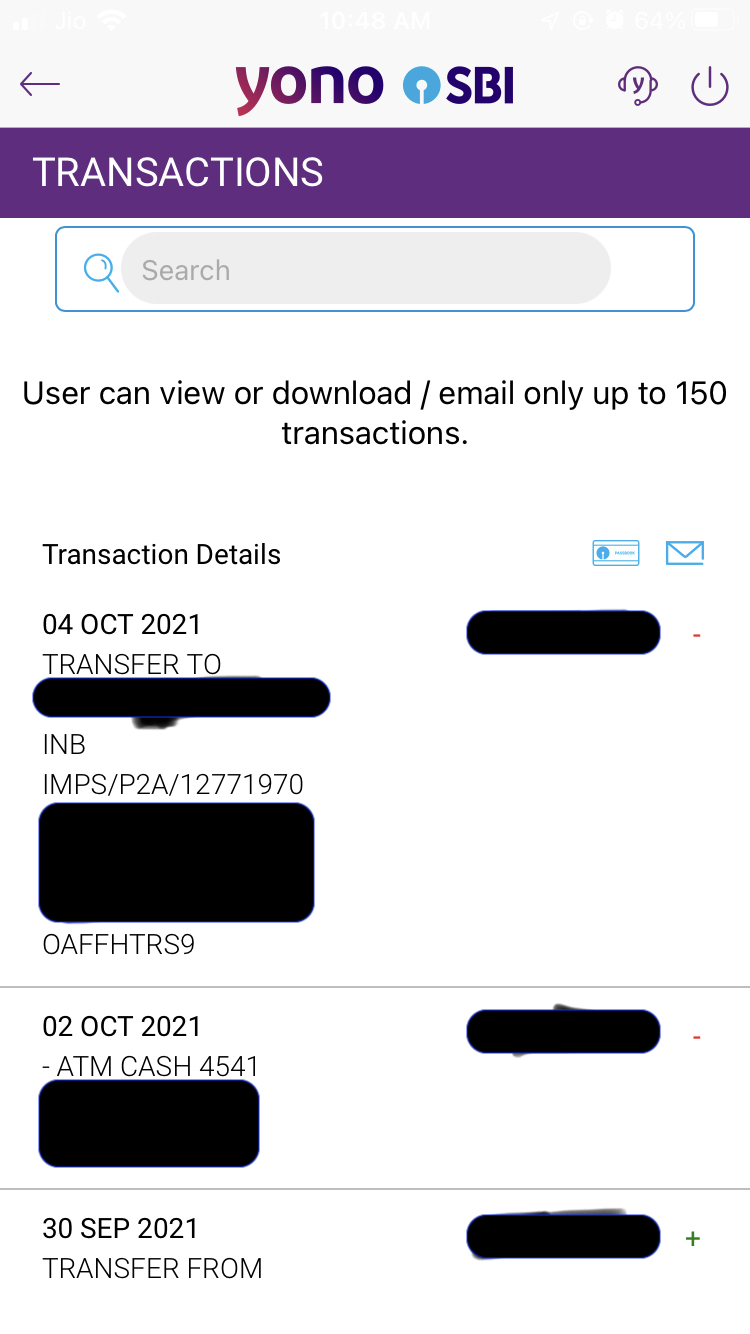}}
  \caption{A banking app interface with potential accessibility issues. Certain details blocked for privacy.}\label{fig:midterm}
  \Description{A banking app interface with potential accessibility issues. Certain details blocked for privacy.}
\end{figure}

The midterm exam presented the students with a screenshot of a banking app (Fig.~\ref{fig:midterm}) and asked them to ``identify at least three UI elements that are likely to have accessibility issues and explain how you will fix those issues''.

Learning objectives:
\begin{itemize}
  \item LO1: \underline{Remember} the accessibility guidelines for mobile apps.
  \item LO2: \underline{Analyze} the given real-life example app for potential accessibility issues.
  \item LO3: \underline{Create} a more accessible version of a given software.
\end{itemize}

The students were expected to identify these issues:

\begin{itemize}
  \item Color contrast in the search bar seems very low (can be checked with \as{} and fixed by setting it to the recommended $4.5:1$),
  \item Size of interactive widgets like the `cheque' or `email' button is possibly smaller than the required $48\times 48$ dp (can be tested using \as{} and fixed by increasing the widget size in the XML specification of the UI), and
  \item The interactive widgets, such as the `search button' may be missing appropriate labels (can be tested using \tb{} and fixed by adding the XML attribute in the UI).
\end{itemize}

\subsubsection{Final exam question} The final exam question presented an XML code representing the UI with two input text boxes for the title and author of a book and a clickable button to add this information to the book database and posed these two subquestions:

\begin{enumerate}
  \item Suppose you run this app through \tb{}. Name \emph{two} possible accessibility issues it \emph{will} highlight.
  \item Name \emph{one} accessibility issue \tb{} \emph{will not} highlight.
\end{enumerate}

Learning objectives:
\begin{itemize}
  \item LO4: \underline{Remember} the accessibility guidelines for mobile apps.
  \item LO5: \underline{Evaluate} (critique) an assistive technology tool (\tb).
\end{itemize}

The expected answer to the first subquestion, i.e., the two problems \tb{} will help catch are: 
(a) Since the XML code for the input boxes do not have a hint-text or content-description, \tb{} cannot give a meaningful information about these to the user, and 
(b) Content-description of the `Add Book' button says ``Click to add book''. \tb{} already adds ``Click'' when announcing interactive buttons. So it will say ``Click Click to add book'' which is confusing for a visually impaired user.

The XML code did not specify dimensions for the clickable widget and students had experienced in class that the default size is smaller than the recommended $48\times 48$ dp. Similarly, the default color contrast for the button labels is less than the recommended $4.5:1$. However, \tb{} is not able to catch these problems. For the second subquestion, students were expected to identify either of these two issues.

Both exams were open-book, open-internet and conducted online via GitHub Classroom. 

\subsubsection{Findings}


26 students identified all three accessibility issues in the midterm exam, while the remaining 24 identified one or two of the issues ($N=50$). Forty students proposed a valid solution to the issue(s) they identified.

For the first subquestion on the final exam, 49 participants correctly identified that \tb{} can highlight the missing hint text or content descriptions on the input boxes. But only one student was able to identify the issue with inappropriate content description. 18 students have misidentified that \tb{} will highlight low color contrast and small widget sizes. 

For the second subquestion, 32 students correctly answered that \tb{} cannot identify the issue with small touch targets and 23 answered that it cannot identify the issue with low color contrast (a total of 42 unique students answered this correctly). Nine of them mentioned \as{} can catch these problems. However, 11 students misidentified that \tb{} \emph{cannot} identify the improper content description.
\section{Discussion}
\label{sec:discussion}


\subsection{Accessibility Awareness}

We used the pre-post questionnaire to assess accessibility awareness in the participants. 
As expected, many participants had taken into account infrastructure barriers and diversity due to age, literacy, etc., when designing the hypothetical COVID-19 vaccine verification app at the start of the course. This is due to the fact that students are exposed to such diversity in their daily lives. 
Only one student had considered users with disabilities while designing or testing the solution at the start of the course. Sixteen other students mentioned this after completing the course, confirming that a significant change in awareness about accessibility considerations in software development has been achieved.

A rise in motivation to consider accessibility is also visible in Assignment 4.
Over a third of the participants ($9+8=17$ out of 50) demonstrated intrinsic goal orientation by either performing accessibility tests or expressing their plans to do so on the last assignment even when it was not a stated requirement.
Intrinsic or mastery goal orientation is ``the degree to which a student values learning for the sake of personal growth'' (as against for the sake of other factors such as grades) and is generally associated with ``greater success in academic situations''~\citep{lishinskiMotivationAttitudesDispositions2019}.


\subsection{Technical Knowledge}

Instruments $B$ and $C$, the programming assignments and exam questions, evaluate the technical knowledge of students about accessibility implementation and testing. These instruments have accessibility related learning objectives that cover all six levels of the revised Bloom's Taxonomy (Sec.~\ref{sec:method:assignments} and \ref{sec:method:exams}).
Our findings demonstrate that the interventions helped students develop their technical knowledge related to implementing and testing accessibility in mobile app development.


On the midterm question, over half of the students identified all three accessibility issues by looking at the UI of a popular banking app, and the remaining half identified at least one issue. 90\% of the students were able to suggest a valid fix to the accessibility issues they found.

On the final exam question, although nearly everyone correctly pointed out that \tb{} can be used to catch the problem with missing hint-text or content description, only one student pointed out its use to catch the inappropriate label where verbs like `click' will be repeated. 
This is most likely because although this use of \tb{} was discussed in class, the examples in class exposed them only to missing labels and not to such `inappropriate' labels. 
Final exam answers also revealed that about two-fifths of the students have misconceptions about the abilities and limitations of the assistive technology tool \tb{}.


\subsection{Empathy}

Although we do not have an instrument to measure empathy, responses to reflective questions on assignments 1 and 2 reveal that after using \tb{} blindfolded, many students empathized with the difficulties faced by persons with disabilities (PWDs) in using software that is not accessible. 
For instance, a participant stated that s/he ``...realized why [this exercise] is important as using \tb{} people with disabilities can also use the applications.''
Since doing this exercise in A1, some students started using appropriate accessibility best practices proactively and reported the ease with which they could now use \tb: ``Using \tb{} was a fun experience this time as I had added contentDescription for the various views before doing this test.'' 
Similarly, another student stated she started adding appropriate contentDescriptions wherever required to ensure that ``my app will be accessible for visually impaired or people with weaker eyesight as well.''
This is consistent with existing literature that suggests such empathy helps students understand the importance of developing accessible software~\citep{el-glalyPresentingEvaluatingImpact2020}.

\section{Limitations and Future Work}
\label{sec:future}

The accessibility topics in this offering of the course covered visual impairments. In future iterations of the course, other disabilities like hearing, motor, and cognitive impairments should also be considered.

This work needs to be replicated in varied contexts: ours was a class of junior/senior undergraduate students in an engineering institute, but mobile app development is taught in a variety of computing education programs at undergrad and graduate levels (e.g., a Bachelor of Computer Science or Master of Computer Applications program in India). It will be interesting to see if our results remain consistent across such variations.

Since most of our students learn Java in their sophomore year, the course is designed to use Java for Android programming. However, there is a growing demand for moving to Kotlin and covering cross-platform development using React Native and Flutter. New course material should be developed to cover accessibility topics using these platforms.

Interacting with PWDs can substantially impact student empathy and motivation to build accessible software~\citep{zhaoComparisonMethodsTeaching2020}. Since this course iteration was online due to the pandemic, it was impossible to include PWDs in the design or testing of the assignments. In a future offering, we will like to arrange for such interactions between students and PWDs and study their effects on student learning.

It should be noted that while the exam data are actual student answers, the assignment data are self-reported responses to reflective questions. Analyzing actual student code to see how they have implemented and tested accessibility features can reveal some interesting findings.

We also plan to interview students who considered accessibility in the last assignment even though it was not a stated requirement and those who did not consider accessibility in other assignments to understand the challenges they possibly faced.

\section{Conclusion}
\label{sec:conclusion}

This work reports on our experience integrating accessibility in a regular mobile app development course that uses Java-based Android programming. Our findings demonstrate that: 
(a) students' awareness about considering accessibility when designing, developing, and testing software has increased after completing the course, 
(b) students acquired technical knowledge about accessibility guidelines, tools to test for accessibility features, and best practices for implementing accessibility in Android apps, and 
(c) empathy-creating exercises like interacting with the mobile device blindfolded using a screen reader resulted in increased empathy for challenges faced by persons with disabilities when using inaccessible software. 
Other educators can teach accessibility topics in similar courses without diluting their core learning objectives and without much prior knowledge about accessibility.
All course material and resources are publicly available at \url{https://swaroopjoshi.in/project/sugamyata/}.

\begin{acks}
  This work is supported by Birla Institute of Technology and Science, Pilani under grants BPGC/RIG/2020-21/04-2021/02 and GOA/ACG/2021-2022/Nov/05.
\end{acks}

\bibliographystyle{ACM-Reference-Format}




\end{document}